\documentclass[11pt]{article}
\usepackage[numbers]{natbib}
\usepackage{a4}
\usepackage{citesort}
\usepackage{amsmath}
\usepackage{amssymb}
\usepackage{latexsym}
\usepackage{amssymb}
\usepackage{epsfig}
\usepackage{natbib}
\def \TT{{\mathrm{T}}}
\def \TL{{\mathrm{L}}}

\def \d{{\mathrm{d}}}

\def \R{{\mathbb{R}}}

\def \pd{\partial}

\def \Bsigma{\boldsymbol{\sigma}}
\def \Bbeta{\boldsymbol{\beta}}

\def \Bs{\boldsymbol{s}}

\def \Bu{{\boldsymbol{u}}}

\def \Bv{{\boldsymbol{v}}}
\def \BR{{\boldsymbol{R}}}
\def \BV{{\boldsymbol{V}}}
\def \BQ{{\boldsymbol{Q}}}
\def \BF{{\boldsymbol{F}}}

\def \Bbeta{\boldsymbol{\beta}}

\def \rr{{\boldsymbol{r}}}
\def \RR{{\boldsymbol{R}}}

\def \Bu{{\boldsymbol{u}}}
\def \Bv{{\boldsymbol{v}}}

\def \Bp{{\boldsymbol{p}}}

\begin{document}
\title{{\bf 
The elastodynamic Li\'enard-Wiechert potentials and elastic fields 
of non-uniformly moving point and line forces
}}
\author{
Markus Lazar~$^\text{a,b,}$\footnote{
{\it E-mail address:} lazar@fkp.tu-darmstadt.de (M.~Lazar).
}
\\ \\
${}^\text{a}$ 
        Heisenberg Research Group,\\
        Department of Physics,\\
        Darmstadt University of Technology,\\
        Hochschulstr. 6,\\      
        D-64289 Darmstadt, Germany\\
${}^\text{b}$ 
Department of Physics,\\
Michigan Technological University,\\
Houghton, MI 49931, USA
}

\date{\today}    
\maketitle

\begin{abstract}
The purpose of this paper is to investigate 
the fundamental problem of the non-uniform subsonic motion of a point force 
and line forces in an unbounded, homogeneous, isotropic medium
in analogy to the electromagnetic  Li\'enard-Wiechert potentials.
The exact closed-form solutions of the 
displacement and elastic fields 
produced by the point force and line forces are  calculated. 
The displacement fields can be identified with the elastodynamic
Li\'enard-Wiechert tensor potentials. 
For a non-uniformly moving point force, 
we decompose the elastic fields into a radiation part and a 
non-radiation part. 
We show that the solution of a non-uniformly moving point force is the  
generalization of the Stokes solution towards the non-uniform motion.
For line forces the mathematical solutions are given in the form of 
time-integrals and, therefore, their motion depends on the history.
\\

\end{abstract}
\newpage

\section{Introduction}

An important item in elastodynamics is
concerned with the radiation and the 
waves produced by the non-uniform motion of 
body forces.
This is a fascinating and interdisciplinary research topic. 
The radiation problem 
has attracted the interest of researchers from different fields such as
applied mathematics, material science, continuum mechanics, and 
seismology (see, e.g, \citep{Achenbach,Miklowitz,Hudson,AR,Pujol}).
A fundamental question is: what is the elastic radiation caused 
by non-uniformly moving point forces?

In elastostatics, the so-called Kelvin problem concerns with the 
displacement and elastic fields produced by a static point force.
In elastodynamics the displacement field generated by a  
time-dependent concentrated point load was first presented by~\citet{Stokes} 
(see, e.g., \citep{Achenbach,AR,Gurtin}).
In the Stokes problem, the body force is considered as a concentrated load of
time-dependent magnitude.
The Stokes solution can be considered as 
the first mathematical model of an earthquake~\citep{BS}.
Concentrated line forces 
with time-dependent magnitude were studied by~\citet{deHoop58} 
and \citet{Achenbach}.
The wave-motion caused by a line force moving non-uniformly in a fixed
direction was considered by~\citet{Freund72}.
A non-uniformly moving line force in an anisotropic elastic solid was studied 
by~\citet{Wu02}.

The radiation problem of point forces is three-dimensional so that 
Huygens' principle prevails.
Using the Helmholtz decomposition,
the so-called retarded potentials were given
for the waves produced by body forces in elastodynamics
(see, e.g., \citep{Achenbach,Miklowitz}).
A more general expression for the retarded potential in elastodynamics was given 
by~\citet{Hudson}.
Elastodynamic fields propagate with finite velocities.
There always is a time-delay before a change in elastodynamic conditions 
initiated at a point of space can produce an effect at any other point of
space. This time-delay is called elastodynamic retardation.

In electrodynamics,
radiation is caused by the 
non-uniform motion of an electric point charge. 
The electric and magnetic potentials of such a non-uniformly moving point
charge are called the Li\'enard-Wiechert potentials.
The corresponding electric and magnetic field strengths 
consist of velocity-depending fields and acceleration-depending fields.
The last ones are the fields of radiation.
This is a standard topic in electromagnetic field theory and is
covered in a lot of books on electrodynamics (e.g. \citep{LL,Jackson}).
It is quite surprising that nothing has been investigated
in this direction in the elastodynamics of moving point forces.
No solution of a non-uniformly moving point force 
analogous to the Li\'enard-Wiechert potential can be found in standard books 
on elastic waves (e.g.~\citep{Achenbach,Miklowitz,Eringen75,deHoop,Hudson,AR,Pujol}).

The purpose of the present paper is to investigate 
the fundamental problem of the non-uniform motion of a point force 
as well as line forces in 
an unbounded, homogeneous, isotropic medium
in analogy to the electromagnetic  Li\'enard-Wiechert potentials.
We consider the subsonic motion.
The paper is organized as follows.
In Section~2, we present 
the framework of elastodynamics and we formulate
the equation of motion.
In Section~3, using the three-dimensional elastodynamic 
Green tensor, we calculate the elastodynamic 
Li\'enard-Wiechert potential of a point force.
In Section~4, using the 
Li\'enard-Wiechert potential of a point force, we determine the 
elastic distortion and the velocity fields (particle velocity)
of the medium caused by the non-uniformly moving point force. 
In addition, we specify the radiation fields proportional to the
acceleration of the point force.
The limit to the Stokes solution is performed in Section~5.
The static limit of the displacement and elastic fields 
of the non-uniformly moving point force is given in Section~6.
In Section~7, using the two-dimensional Green tensors, 
we give the general solution of the two-dimensional 
non-uniformly moving line forces.
We close the paper with conclusions in Section~8.

\section{The elastodynamic equation of motion}

In elastodynamics~\citep{Gurtin}, the force balance law reads\footnote{Spatial
differentiation is denoted by $\pd_j\equiv \pd/\pd x_j$, and 
for the differentiation with respect to time $t$ we 
use the notation $\dot{p}_{i}\equiv\pd_t p_{i}$.}
\begin{align}
\label{EC0}
\dot{p}_i -\pd_j \sigma_{ij}=F_i\,,
\end{align}
where $\Bp$, $\Bsigma$ and $\BF$ are the linear momentum vector, the force
stress tensor and the body force vector. 
In the theory of linear elasticity, 
the momentum vector $\Bp$ and the stress tensor 
$\Bsigma$ can be expressed in terms of the
physical state quantities, namely, 
the velocity vector (particle velocity) $\Bv=\dot{\Bu}$ 
and the elastic distortion tensor $\Bbeta=({\text{grad}}\,\Bu)^\TT$ of the medium
which can be derived from a displacement vector $\Bu$
by means of the following constitutive relations
\begin{align}
\label{CR-p}
p_i&= \rho\,  v_i=\rho\, \dot{u}_i\,,\\
\label{CR-t}
\sigma_{ij}&=C_{ijkl}\, \beta_{kl}=C_{ijkl}\, \pd_l u_{k}\,,
\end{align}
where $\rho$ denotes the mass density and 
$C_{ijkl}$ is the tensor of elastic moduli.
The tensor $C_{ijkl}$ possesses the following symmetry properties
\begin{align}
C_{ijkl}=C_{jikl}=C_{ijlk}=C_{klij}\, .
\end{align}
If we substitute the constitutive relations~(\ref{CR-p}) and (\ref{CR-t}) in 
Eq.~(\ref{EC0}), we obtain the force balance law expressed in terms of 
the displacement vector $\Bu$ 
\begin{align}
\label{EC}
\rho\, \ddot{u}_i -C_{ijkl}\pd_j\pd_l u_{k}=F_i\, .
\end{align}

The solution of Eq. (\ref{EC}) can be represented as a convolution-integral
in space and time.
In an unbounded medium and under the assumption of zero initial conditions,
which means that $\Bu(\rr,t_0)$ and $\dot\Bu(\rr,t_0)$ are zero 
for $t_0\rightarrow-\infty$, 
the solution of $\Bu$ reads
\begin{align}
u_i(\rr,t)=
\label{u-M}
\int_{-\infty}^t \int_{-\infty}^\infty
G_{ij}(\rr-\rr', t-t')\, F_{j}(\rr',t')\, \d \rr'\, \d t'\, .
\end{align}
Here, $G_{ij}$ is the elastodynamic Green tensor of the anisotropic Navier equation 
defined by
\begin{align}
\label{GF-e}
\big[\delta_{ik}\,\rho \, \pd_{tt}-C_{ijkl}\pd_j\pd_l\big] G_{km}=\delta_{im}\,
\delta(t)\delta(\rr)\,,
\end{align}
where $\delta(.)$ denotes the Dirac delta function and 
$\delta_{ij}$ is the Kronecker delta.
The tensor of elastic moduli for isotropic materials is given by
\begin{align}
\label{C}
C_{ijkl}=\lambda\, \delta_{ij}\delta_{kl}
+\mu\big(\delta_{ik}\delta_{jl}+\delta_{il}\delta_{jk})\,,
\end{align}
where $\lambda$ and $\mu$ are the Lam{\'e} constants.
Substituting Eq.~(\ref{C}) in Eqs.~(\ref{EC}) and (\ref{GF-e}), we 
obtain respectively 
the isotropic Navier equations for the displacement vector
\begin{align}
\label{NE-iso}
\big[\delta_{ij}\,\rho \, \pd_{tt}- \delta_{ij}\, \mu\, \Delta
-(\lambda+\mu)\, \pd_i \pd_j \big] u_{j}=F_i\, ,
\end{align}
and for the elastodynamic Green tensor
\begin{align}
\big[\delta_{ij}\,\rho \, \pd_{tt}- \delta_{ij}\, \mu\, \Delta
-(\lambda+\mu)\, \pd_i \pd_j \big] G_{jm}=\delta_{im}\,
\delta(t)\delta(\rr)\,,
\end{align}
where $\Delta$ denotes the Laplacian.

When the material is isotropic and infinitely extended, 
the three-dimensional elastodynamic Green tensor reads~\citep{Love,Eringen75,AR,Pujol}
\begin{align}
\label{GT}
G_{ij}(\rr,t)&=\frac{1}{4\pi\rho }\, 
\Bigg\{
\frac{\delta_{ij}}{r c^2_\TT}\, \delta(t-r/c_\TT)
+\frac{x_i x_j}{r^3}\,
\bigg(
\frac{1}{c^2_\TL}\,  \delta(t-r/c_\TL)
-\frac{1}{c^2_\TT}\,  \delta(t-r/c_\TT)\bigg)\nonumber\\
&\hspace{15mm}
+\bigg(\frac{3x_i x_j}{r^2}-\delta_{ij}\bigg)
\frac{1}{r^3}\,\int_{r/c_\TL}^{r/c_\TT}\tau\,\delta(t-\tau)\,\d \tau
\Bigg\}\,,
\end{align}
where $r=\sqrt{x_1^2+x_2^2+x_3^2}$.
It should be pointed out that the tensor 
in Eq.~(\ref{GT}) is the retarded Green tensor.
Here $c_\TL$ and $c_\TT$ denote the velocities of the 
longitudinal and transversal elastic waves (sometimes called P- and S-waves). 
The two sound-velocities can be given in terms of the Lam\'e constants ($c_\TT<c_\TL$)
\begin{align}
\label{c}
c_{\TL}=\sqrt{\frac{2\mu+\lambda}{\rho}}\,,\qquad
c_{\TT}=\sqrt{\frac{\mu}{\rho}}\, .
\end{align}
The elastodynamic Green tensor~(\ref{GT}) is a tensor with support
along the two sound-cones $r=c_\TT t$ and $r=c_\TL t$ 
as well as in between them. 
Moreover, it should be mentioned that
it consists of near-field and far-field terms.
The first two terms in Eq.~(\ref{GT}) decay as $1/r$ and, thus, 
they are the far-field terms. 
The last term in Eq.~(\ref{GT}) decays more rapidly as $1/r^2$ which gives 
the near-field term (see, e.g., \citep{Chapman}).

\section{The elastodynamic Li\'enard-Wiechert potential of a point force}
Now we consider the non-uniform motion of a point force
of total strength $Q_j(t)$, situated at the position $\Bs(t)$.
Then the point force is
\begin{align}
\label{F}
F_{i}(\rr,t)&=Q_i(t) \, \delta(\rr-\Bs(t))\,
\qquad
\text{for}\ \rr\in\R^3,\ t\in\R\,.
\end{align}
Moreover, only subsonic source-speeds will be admitted ($|\BV|<c_\TT$).
Substitution of Eq.~(\ref{F}) in Eq.~(\ref{u-M}) 
and integration in $\rr'$ lead to
\begin{align}
u_i(\rr,t)=
\label{u-M-L}
\int_{-\infty}^t 
 G_{ij}(\rr-\Bs(t'), t-t')\, Q_j(t')\, \d t'\, .
\end{align}
The structure of the Green tensor~(\ref{GT}) 
produces in Eq.~(\ref{u-M-L}) three characteristic integrals 
which we have to calculate. 
If we substitute the elastodynamical Green tensor (\ref{GT}) 
in Eq.~(\ref{u-M-L}) and use the relation
\begin{align}
\label{Rel}
\frac{1}{r^2}\,\int_{r/c_\TL}^{r/c_\TT}\tau\,\delta(t-\tau)\,\d \tau=
\int_{1/c_\TL}^{1/c_\TT}\kappa\,\delta(t-\kappa r)\,\d \kappa\,,
\end{align}
where $\kappa$ is a dummy variable with the dimension of slowness $=1/[\text{velocity}]$,
we obtain
\begin{align}
u_i(\rr,t)&=\frac{1}{4\pi\rho }
\label{u-M-L-2}
\int_{-\infty}^t 
\Bigg\{
\frac{\delta_{ij}\, \delta(t-t'-R(t')/c_\TT)}{c^2_\TT\, R(t')}
\nonumber\\
&\quad
+\frac{R_i(t') R_j(t')}{R^3(t')}\,
\bigg(
\frac{1}{c^2_\TL}\,  
\delta\big(t-t'-R(t')/c_\TL\big)
-\frac{1}{c^2_\TT}\,  
\delta\big(t-t'-R(t')/c_\TT\big)\bigg)\nonumber\\
&\quad
+\bigg(\frac{3R_i(t') R_j(t')}{R^3(t')}
-\frac{\delta_{ij}}{R(t')}\bigg)
\int_{1/c_\TL}^{1/c_\TT}
\kappa\,\delta(t-t'-\kappa\, R(t'))\,\d \kappa
\Bigg\}
\, Q_j(t')\, \d t'\, ,
\end{align}
where $\BR(t')=\rr-\Bs(t')$ and $R(t')=[R_m(t')R_m(t')]^{1/2}=|\rr-\Bs(t')|$. 
Now the time-integration in Eq.~(\ref{u-M-L-2}) can be performed. 
We express the integrals in terms of retarded variables by appeal to the 
relation~\citep{Roos,Barton}
\begin{align}
\label{Roos}
\int \delta(f(t'))\,g(t')\, \d t'=\frac{g(t')}{|\d f/\d t'|}\bigg|_{{\text{at}}\, f(t')=0}
\, .
\end{align}
Mathematically, the factor $1/|\d f/\d t'|$ is the Jacobian of the 
transformation from $t'$ to the new integration variable $f(t')$.
This mapping between the two variables is one-to-one if the Jacobian is different from zero. A sufficient condition for this is that the velocity of the source 
(point force) is less than the slowest wave speed ($|\BV|<c_\TT$).
The first integral of Eq.~(\ref{u-M-L-2}) can be carried out with
\begin{align}
\label{Int1}
\int \frac{\delta(t-t'-R(t')/c_\TT)\, Q_j(t')}{R(t')}\, \d t'
=\frac{Q_j(t')}{R(t')-V_m(t') R_m(t')/c_\TT}\bigg|_{t'=t_\TT}\,,
\end{align}
where we introduced the so-called transversal retarded time $t_\TT=t'(\rr,t)$
which is the solution of the condition
\begin{align}
\label{tT}
t-t'-R(t')/c_\TT=0\,.
\end{align}
$\BV=\dot{\Bs}$ is the velocity of the moving body force and
$\BR$ is the distance vector 
from the position of the point force $\Bs$, the sender of
elastic waves, to the point of the observer $\rr$, the receiver of the
elastic waves.
The second integral is
\begin{align}
\label{Int2}
\int \frac{R_i(t') R_j(t')\, Q_j(t')}{R^3(t')}\, 
\delta(t-t'-R(t')/c_{\TL,\TT})\, \d t'
=\frac{R_i(t')R_j(t')}{R^2(t')}\,\frac{Q_j(t')}{R(t')- V_m(t') R_m(t')/c_{\TL,\TT}}\bigg|_{t'=t_{\TL,\TT}}\, .
\end{align}
Here $t_\TL=t'(\rr,t)$ denotes the longitudinal retarded time which is the solution
of the equation
\begin{align}
\label{tL}
t-t'-R(t')/c_\TL=0\,.
\end{align}
We perform the third integral as follows
\begin{align}
\label{Int3}
&\int \bigg(\frac{3R_i(t') R_j(t')}{R^3(t')}
-\frac{\delta_{ij}}{R(t')}\bigg) Q_j(t')
\int_{1/c_\TL}^{1/c_\TT}\kappa\,
\delta(t-t'-\kappa\, R(t'))\,\d \kappa\, \d t'
\nonumber\\
&\qquad\quad
=\int_{1/c_\TL}^{1/c_\TT}
\bigg(\frac{3R_i(t') R_j(t')}{R^2(t')}-\delta_{ij}\bigg)
\frac{ Q_j(t')\, \kappa\,\d \kappa}{R(t')-\kappa\, V_m(t')R_m(t')}\bigg|_{t'=t_\kappa}\, 
\end{align}
with $t_\kappa=t'(\rr,t)$ as solution of the equation
\begin{align}
\label{tlambda}
t-t'-\kappa\, R(t')=0\, .
\end{align}
The retarded time $t_\kappa$ is an effective retarded time for the 
$\kappa$-integration with limits $(1/c_\TL,1/c_\TT)$.
The solutions of $t_\TT$, $t_\TL$, $t_\kappa$ are unique if 
$|\BV|$ is less than $c_\TT$. 
Thus, for subsonic motion the solutions of Eqs.~(\ref{tT}), (\ref{tL}) and 
(\ref{tlambda}) are unique.
The retarded times are a result of the finite speeds of 
propagation for elastodynamic waves.
In Eqs.~(\ref{Int1}), (\ref{Int2}) and (\ref{Int3}) we have used the relation
\begin{align}
\bigg|\frac{\d f(t')}{\d t'}\bigg|_{t'=t_{\text{ret}}}=
1-\frac{V_m(t') R_m(t')}{c\, R(t')}\bigg|_{t'=t_{\text{ret}}}>0\quad
\text{for}\ |\BV|<c_\TT\,, \ \ c=c_\TT, c_\TL, 1/\kappa\quad
{\text{and}}\  t_{\text{ret}}=t_\TT, t_\TL, t_\kappa\,,
\end{align}
where $f(t')=t-t'- R(t')/c$.

Thus, carrying out the integration in $t'$ in Eq.~(\ref{u-M-L-2}),
we find the explicit expression for the 
displacement field of a non-uniformly moving point force which we call the 
elastodynamic 
Li\'enard-Wiechert potential of a point force
\begin{align}
\label{u}
u_i(\rr,t)&=
\frac{1}{4\pi\rho}
\Bigg\{
\frac{1}{c^2_\TT}
\bigg[\bigg(\delta_{ij}-\frac{R_i(t')R_j(t')}{R^2(t')}\bigg)
 \frac{Q_j(t')}{R(t')-V_m(t') R_m(t')/c_\TT}\bigg]\bigg|_{t'=t_\TT}
\nonumber\\
&\qquad\quad
+\frac{1}{c^2_\TL}\,
\bigg[\frac{R_i(t')R_j(t')}{R^2(t')}\,
 \frac{Q_j(t')}{R(t')-V_m(t') R_m(t')/c_\TL}\bigg]\bigg|_{t'=t_\TL}
\nonumber\\
&\qquad\quad
+\int_{1/c_\TL}^{1/c_\TT}\d \kappa \, \kappa
\bigg[\bigg(\frac{3R_i(t') R_j(t')}{R^2(t')}-\delta_{ij}\bigg)
\frac{Q_j(t')}{R(t')-\kappa\, V_m(t') R_m(t')}\bigg]\bigg|_{t'=t_\kappa}
\Bigg\}\,, 
\end{align}
where $\BR(t')$, $\BV(t')$ and $\BQ(t')$ 
are to be evaluated at the corresponding retarded times.
The first $\delta_{ij}$-term in Eq.~(\ref{u}) has the form 
of the acoustic Li\'enard-Wiechert potential~\citep{Barton,deHoop05}
and if $\BQ$ is time-independent, 
this term reduces to the form of the well-known electric 
Li\'enard-Wiechert potential of a point charge 
in electrodynamics (see, e.g., \citep{LL,Jackson}).
Due to the appearance of two velocities of the elastic waves, 
the elastodynamic Li\'enard-Wiechert potential of a point force 
has a more complicated but rather straightforward structure.
Eq.~(\ref{u}) consists of 
three characteristic pieces. The first term is the transversal one, 
transmitting with speed $c_\TT$, and it corresponds to $S$-wave motion.
The second term is the longitudinal one, 
transmitting with speed $c_\TL$, and it corresponds to $P$-wave motion.
The third term is 
neither longitudinal nor transversal and it gives 
contribution arriving at speeds between the two characteristic ones.
This shows that this factor 
represents a combination of $P$-wave and $S$-wave motion.

It is also important to note that the following elastodynamic Doppler factors appear in Eq.~(\ref{u}):
\begin{align}
&1-n_i(t_{\text{ret}})V_i(t_{\text{ret}})/c\,\quad
\text{for}\ |\BV|<c_\TT\,,\qquad c=c_\TT, c_\TL, 1/\kappa\quad
{\text{and}}\  t_{\text{ret}}=t_\TT, t_\TL, t_\kappa\,,
\end{align}
where $n_i=R_i/R$.

\section{Radiation of the elastic fields of a point force}
In order to derive the elastic fields,
we have to calculate the gradient and the time-derivative of the 
displacement field~(\ref{u}). 
The appearance of the retarded times~(\ref{tT}), (\ref{tL}) and (\ref{tlambda}) in
Eq.~(\ref{u}) makes the differentiation more difficult.
To evaluate these derivatives we need some more basic 
derivatives (see also \citep{Barton}).
We introduce
\begin{align}
\label{P}
P_c(t')=R(t')-V_m(t')R_m(t')/c \,, \qquad c=c_\TT,c_\TL,1/\kappa\,
\end{align}
and obtain after a straightforward calculation
\begin{align}
\label{grad-t}
 \bigg[\frac{\pd t'}{\pd x_k}\bigg]\bigg|_{t'=t_{\text{ret}}}
=-\bigg[\frac{R_k(t')}{c P_c(t')}\bigg]\bigg|_{t'=t_{\text{ret}}}
\end{align}
\begin{align}
\label{grad-Q}
\pd_k \big[Q_j(t')\big]\big|_{t'=t_{\text{ret}}}
=\bigg[\frac{\pd t'}{\pd x_k}\, \frac{\pd Q_j(t')}{\pd t'}\bigg]\bigg|_{t'=t_{\text{ret}}}
=-\bigg[\frac{R_k(t')}{c P_c(t')}\, \dot{Q}_j(t')\bigg]\bigg|_{t'=t_{\text{ret}}}
\end{align}
\begin{align}
\label{div-P}
\pd_k \bigg[\frac{1}{P_c(t')}\bigg]\bigg|_{t'=t_{\text{ret}}}
=-\frac{1}{P^3_c(t')}\bigg[
\frac{\dot{V}_m(t') R_m(t')}{c^2}\, R_k(t')
+\bigg(1-\frac{V^2(t')}{c^2}\bigg)R_k(t')
- \frac{P_c(t')}{c}\, V_k(t')\bigg]\bigg|_{t'=t_{\text{ret}}}
\end{align}
and
\begin{align}
\label{div-R}
\pd_k \bigg[\frac{R_i(t') R_j(t')}{R^2(t')}\bigg]\bigg|_{t'=t_{\text{ret}}}
&=\bigg[
\frac{R_j(t')}{R^2(t') }\bigg(\delta_{ik}+\frac{V_i(t')R_k(t')}{c P_c(t')}\bigg)
+\frac{R_i(t')}{R^2(t')}\bigg(\delta_{jk}+\frac{V_j(t')R_k(t')}{c P_c(t')}\bigg)
\nonumber\\
&\qquad\quad
-\frac{2 R_i(t') R_j(t') R_k(t')}{R^3(t') P_c(t')}\bigg]\bigg|_{t'=t_\text{ret}}\,,
\end{align}
where $\BR(t')$, $R(t')$, $\BV(t')$, $P_c(t')$ and $\BQ(t')$ 
are to be evaluated at the corresponding retarded times.
Using Eqs.~(\ref{grad-Q})--(\ref{div-R}), we find for
the gradient of the displacement field~(\ref{u})
\begin{align}
\label{B}
&\beta_{ik}(\rr,t)=-\frac{1}{4\pi\rho}
\Bigg\{
\frac{1}{c^3_\TT}
\bigg[
\bigg(\delta_{ij}-\frac{R_i(t')R_j(t')}{R^2(t')}\bigg)
 \frac{R_k(t') \dot{Q}_j(t')}{P^2_\TT(t')}\bigg]\bigg|_{t'=t_\TT}
+\frac{1}{c^3_\TL}
\bigg[
\frac{R_i(t')R_j(t')R_k(t')}{R^2(t')}\,
 \frac{\dot{Q}_j(t')}{P^2_\TL(t')}\bigg]\bigg|_{t'=t_\TL}
\nonumber\\
&\qquad\qquad\qquad\qquad
+\int_{1/c_\TL}^{1/c_\TT}\d \kappa\, \kappa^2
\bigg[\bigg(\frac{3R_i(t') R_j(t')}{R^2(t')}-\delta_{ij}\bigg)
\frac{R_k(t')\dot{Q}_j(t')}{P^2_\kappa(t')}\bigg]\bigg|_{t'=t_\kappa}
\nonumber\\
&
+\frac{Q_j(t')}{c^2_\TT}\bigg[
\bigg(\delta_{ij}-\frac{R_i(t')R_j(t')}{R^2(t')}\bigg)
 \frac{1}{P^3_\TT(t')}
\bigg(\frac{\dot{V}_m(t') R_m(t')}{c^2_\TT} R_k(t')
+\bigg(1-\frac{V^2(t')}{c^2_\TT}\bigg)R_k(t')
- \frac{P_\TT(t')}{c_\TT}\, V_k(t')\bigg)
\nonumber\\
&
+\frac{R_j(t')}{R^2(t') P_\TT(t')}
\bigg(\delta_{ik}+\frac{V_i(t')R_k(t')}{c_\TT P_\TT(t')}\bigg)
+\frac{R_i(t')}{R^2(t') P_\TT(t')}
\bigg(\delta_{jk}+\frac{V_j(t')R_k(t')}{c_\TT P_\TT(t')}\bigg)
-\frac{2 R_i(t') R_j(t') R_k(t')}{R^3(t') P^2_\TT(t')}\bigg]\bigg|_{t'=t_\TT}
\nonumber\\
& 
+\frac{Q_j(t')}{c^2_\TL}\bigg[\frac{R_i(t')R_j(t')}{R^2(t')}\, 
\frac{1}{P^3_\TL(t')}
\bigg(\frac{\dot{V}_m(t') R_m(t')}{c^2_\TL}\, R_k(t')
+\bigg(1-\frac{V^2(t')}{c^2_\TL}\bigg)R_k(t')
- \frac{P_\TL(t')}{c_\TL}\, V_k(t')\bigg)
\nonumber\\
&
-\frac{R_j(t')}{R^2(t') P_\TL(t')}
\bigg(\delta_{ik}+\frac{V_i(t')R_k(t')}{c_\TL P_\TL(t')}\bigg)
-\frac{R_i(t')}{R^2(t') P_\TL(t')}
\bigg(\delta_{jk}+\frac{V_j(t')R_k(t')}{c_\TL P_\TL(t')}\bigg)
+\frac{2 R_i(t') R_j(t') R_k(t')}{R^3(t') P^2_\TL(t')}\bigg]\bigg|_{t'=t_\TL}
\nonumber\\
& 
+\int_{1/c_\TL}^{1/c_\TT}
\d \kappa\, \kappa\, Q_j(t') 
\bigg[\bigg(\frac{3R_i(t') R_j(t')}{R^2(t')}-\delta_{ij}\bigg)
\frac{1}{P^3_\kappa(t')}
\bigg(\kappa^2 \dot{V}_m(t') R_m(t') R_k(t')
\nonumber\\
&\hspace{9cm}
+\big(1-\kappa^2 V^2(t')\big)R_k(t')
- \kappa\, P_\kappa(t')  V_k(t')\bigg)
\nonumber\\
&\qquad 
-\frac{3 R_j(t')}{R^2(t') P_\kappa(t')}
\bigg(\delta_{ik}+\frac{\kappa V_i(t')R_k(t')}{P_\kappa(t')}\bigg)
-\frac{3 R_i(t')}{R^2(t') P_\kappa(t')}
\bigg(\delta_{jk}+\frac{\kappa V_j(t')R_k(t')}{P_\kappa(t')}\bigg)
\nonumber\\
&\qquad
+\frac{6 R_i(t') R_j(t') R_k(t')}{R^3(t') P^2_\kappa(t')}\bigg]\bigg|_{t'=t_\kappa}
\Bigg\}\,.
\end{align}
This is the elastic distortion field produced by a non-uniformly moving
point force.
For a constant strength $Q_j=\text{constant}$, 
the $\delta_{ij}$-term in the $t_\TT$-expression agrees with the corresponding
one given in~\citep{Barton} for the gradient of the Li\'enard-Wiechert
potential in acoustics.
The elastic distortion consists of parts depending on the velocity
$\BV$, parts depending on the acceleration $\dot{\BV}$ (elastic radiation part),
and parts depending on $\dot{\BQ}$. 
Note that the dot over $\BV$ and $\BQ$ indicates a derivative with respect to
the argument, namely the corresponding retarded time.

In order to calculate the time derivative of the displacement~(\ref{u}),
we also need the relations
\begin{align}
\label{dt-t}
 \bigg[\frac{\pd t'}{\pd t}\bigg]\bigg|_{t'=t_{\text{ret}}}
=\bigg[\frac{R(t')}{P_c(t')}\bigg]\bigg|_{t'=t_{\text{ret}}}
\end{align}
\begin{align}
\label{dt-Q}
\pd_t \big[Q_j(t')\big]\big|_{t'=t_{\text{ret}}}
=\bigg[\frac{\pd t'}{\pd t}\, \frac{\pd Q_j(t')}{\pd t'}\bigg]\bigg|_{t'=t_{\text{ret}}}
=\bigg[\frac{R(t')}{P_c(t')}\, \dot{Q}_j(t')\bigg]\bigg|_{t'=t_{\text{ret}}}
\end{align}
\begin{align}
\label{dt-P}
\pd_t \bigg[\frac{1}{P_c(t')}\bigg]\bigg|_{t'=t_{\text{ret}}}
=\frac{1}{P^3_c(t')}\bigg[
\big(\dot{V}_m(t') R_m(t') -V^2(t')\big)\, 
\frac{R(t')}{c}+V_m(t') R_m(t')\,\bigg]\bigg|_{t'=t_{\text{ret}}}
\end{align}
and
\begin{align}
\label{dt-R}
\pd_t \bigg[\frac{R_i(t') R_j(t')}{R^2(t')}\bigg]\bigg|_{t'=t_{\text{ret}}}
&=-\bigg[
\frac{1}{R(t') P_c(t')}\big(V_i(t') R_j(t')+V_j(t') R_i(t')\big) 
\nonumber\\
&\qquad\qquad\qquad
-\frac{2 R_i(t') R_j(t') V_m(t') R_m(t')\,}{R^3(t') P_c(t')}
\bigg]\bigg|_{t'=t_\text{ret}}\,.
\end{align}
Using Eqs.~(\ref{dt-Q})--(\ref{dt-R}), we obtain for
the time derivative of the displacement field~(\ref{u})
\begin{align}
\label{v}
&v_{i}(\rr,t)=\frac{1}{4\pi\rho}
\Bigg\{
\frac{1}{c^2_\TT}\bigg[
\bigg(\delta_{ij}R(t')-\frac{R_i(t')R_j(t')}{R(t')}\bigg)
 \frac{\dot{Q}_j(t')}{P^2_\TT(t')}\bigg]\bigg|_{t'=t_\TT}
+\frac{1}{c^2_\TL}
\bigg[
\frac{R_i(t')R_j(t')}{R(t')}\,
 \frac{\dot{Q}_j(t')}{P^2_\TL(t')}\bigg]\bigg|_{t'=t_\TL}
\nonumber\\
&\qquad\qquad\qquad
+\int_{1/c_\TL}^{1/c_\TT}\d\kappa\, \kappa \bigg[
\bigg(\frac{3R_i(t') R_j(t')}{R(t')}-\delta_{ij} R(t')\bigg)
\frac{\dot{Q}_j(t')}{P^2_\kappa(t')}\bigg]\bigg|_{t'=t_\kappa}
\nonumber\\
&\quad
+\frac{Q_j(t')}{c^2_\TT}\bigg[
\bigg(\delta_{ij}-\frac{R_i(t')R_j(t')}{R^2(t')}\bigg) \frac{1}{P^3_\TT(t')}
\bigg(\big[\dot{V}_m(t') R_m(t')-V^2(t')\big]
\frac{R(t')}{c_\TT}+V_m(t') R_m(t')\bigg)
\nonumber\\
&\qquad\qquad\qquad\qquad
+\frac{1}{R(t') P^2_\TT(t')}\big(V_i(t') R_j(t')+V_j(t') R_i(t')\big)
-\frac{2 R_i(t') R_j(t') V_m(t') R_m(t')}{R^3(t') P^2_\TT(t')}\bigg]\bigg|_{t'=t_\TT}
\nonumber\\
&\quad
+\frac{Q_j(t')}{c^2_\TL}\bigg[\frac{R_i(t')R_j(t')}{R^2(t')}
 \frac{1}{P^3_\TL(t')}
\bigg(\big[\dot{V}_m(t') R_m(t')-V^2(t')\big]
\frac{R(t')}{c_\TL}+V_m(t') R_m(t')\bigg)
\nonumber\\
&\qquad\qquad\qquad\qquad
-\frac{1}{R(t') P^2_\TL(t')}\big(V_i(t') R_j(t')+V_j(t') R_i(t')\big)
+\frac{2 R_i(t') R_j(t') V_m(t') R_m(t')}{R^3(t') P^2_\TL(t')}\bigg]\bigg|_{t'=t_\TL}
\nonumber\\
&
+\int_{1/c_\TL}^{1/c_\TT}
\d \kappa\, \kappa\, Q_j(t') 
\bigg[\bigg(\frac{3R_i(t') R_j(t')}{R^2(t')}-\delta_{ij}\bigg)
\frac{1}{P^3_\kappa(t')}
\Big(\big[\dot{V}_m(t') R_m(t') -V^2(t')\big] \kappa R(t') +V_m(t') R_m(t')\Big)
\nonumber\\
&\qquad\qquad\
-\frac{3}{R(t') P^2_\kappa(t')}\big(V_i(t') R_j(t')+V_j(t') R_i(t')\big)
+\frac{6 R_i(t') R_j(t') V_m(t') R_m(t') }{R^3(t') P^2_\kappa(t')}\bigg]\bigg|_{t'=t_\kappa}
\Bigg\}\,.
\end{align}
This is the velocity field (particle velocity) produced by a non-uniformly moving
point force.
For a constant strength $Q_j=\text{constant}$, 
the $\delta_{ij}$-term in the $t_\TT$-expression agrees with the corresponding
one given in~\citep{Barton} for the time derivative of the Li\'enard-Wiechert
potential in acoustics.
Again the velocity vector consists of parts depending on the velocity $\BV$, 
fields depending on the acceleration $\dot{\BV}$ (radiation part),
and fields depending on $\dot{\BQ}$. 
The dot over $\BV$ and $\BQ$ indicates again a derivative with respect to
the argument, namely the corresponding retarded time.

\section{The Stokes solution as limit of a non-uniformly moving point force}
In the current section, we show that the Stokes solution is contained 
in our general results for a non-uniformly moving point force.
In this sense, our solutions~(\ref{u}) and (\ref{B})
are the correct generalizations of the Stokes solution towards the non-uniform motion.
If the position of the point force is fixed, which means that 
$\Bs$ is time-independent and therefore $\BV=0$, 
we recover from the displacement~(\ref{u})
the famous Stokes solution of a concentrated point force with time-dependent
magnitude (e.g.~\citep{Gurtin,Eringen75})
\begin{align}
\label{u-St}
u_i(\rr,t)&=
\frac{1}{4\pi\rho R}\,
\Bigg\{
\frac{1}{c^2_\TT}
\bigg(\delta_{ij}-\frac{R_iR_j}{R^2}\bigg)Q_j(t-R/c_\TT)
+\frac{1}{c^2_\TL}\,
\frac{R_iR_j}{R^2}\,Q_j(t-R/c_\TL)
\nonumber\\
&\qquad
+\bigg(\frac{3R_i R_j}{R^2}-\delta_{ij}\bigg)
\int_{1/c_\TL}^{1/c_\TT}
\kappa\, Q_j(t-\kappa R)\,\d \kappa\Bigg\} \,.
\end{align}
The first terms in Eq.~(\ref{u-St}) 
are usually called the far-field terms since they behave as $1/R$
and the last term in Eq.~(\ref{u-St}) 
is called the near-field term (see, e.g, \citep{AR,Pujol}).
From Eq.~(\ref{B}) and after some mathematical manipulations, 
we find the corresponding displacement gradient
of the Stokes solution (e.g.~\citep{Gurtin,Eringen75})
\begin{align}
\label{B-St}
\beta_{ik}(\rr,t)&=
-\frac{1}{4\pi\rho}
\Bigg\{
3\bigg(\frac{5 R_iR_jR_k}{R^5}
-\frac{\delta_{ij}R_k+ \delta_{jk}R_i+\delta_{ik}R_j}{R^3}\bigg)  
\int_{1/c_\TL}^{1/c_\TT}
\kappa\, Q_j(t-\kappa R)\,\d \kappa\nonumber\\
&\quad
+\bigg(\frac{6R_iR_jR_k}{R^5}
-\frac{\delta_{ij}R_k+ \delta_{jk}R_i+\delta_{ik}R_j}{R^3}\bigg)  
\bigg[
\frac{1}{c^2_\TL}\,Q_j(t-R/c_\TL)
-\frac{1}{c^2_\TT}\,Q_j(t-R/c_\TT)\bigg]\nonumber\\
&\qquad
+\frac{\delta_{ij} R_k}{c^2_\TT R^3}\, 
\bigg[Q_j(t-R/c_\TT)
+\frac{R}{c_\TT}\,\dot{Q}_j(t-R/c_\TT)\bigg]\nonumber\\
&\qquad
+\frac{R_iR_jR_k}{R^4}
\bigg[
\frac{1}{c^3_\TL}\,\dot{Q}_j(t-R/c_\TL)
-\frac{1}{c^3_\TT}\,\dot{Q}_j(t-R/c_\TT)\bigg]
\Bigg\}
 \,.
\end{align}
Thus, Eqs.~(\ref{u}) and (\ref{B}) give the correct Stokes solution
as limit.
It can be seen that the $\dot{\BQ}$-terms in Eq.~(\ref{B-St})
behave as $1/R$ (far-field terms or radiation terms) and
the $\BQ$-terms behave as $1/R^2$ (near-field terms).

\section{Static limit of a non-uniformly moving point force}
In this section, we give the static limit of the Li\'enard-Wiechert
potential and the elastic distortion of a point force as a further check of the
obtained results.

For the static limit, we set
$\BV=0$ and $\BQ=\text{constant}$ and 
substitute Eq.~(\ref{c}) and $\lambda=2\mu\nu/(1-2\nu)$
in Eq.~(\ref{u}). If we perform the integration in $\kappa$ and 
arrange in proper order the appearing terms, we recover 
the displacement field of the Kelvin problem~(see, e.g.,~\citep{Gurtin})
\begin{align}
\label{u-s}
u_{i}(\rr)=\frac{Q_j}{16\pi\mu(1-\nu)R}\bigg[\big(3-4\nu\big)\delta_{ij}
+\frac{R_iR_j}{R^2}\bigg]\,,
\end{align}
where $\RR=\rr-\rr'$ and $\nu$ is Poisson's ratio.
From Eq.~(\ref{B}), we find in the static limit
the displacement gradient of the Kelvin problem~(see, e.g.,~\citep{Gurtin})
\begin{align}
\label{B-s}
\beta_{ik}(\rr)=-\frac{Q_j}{16\pi\mu(1-\nu)R^3}\bigg[\big(3-4\nu\big)\delta_{ij} R_k
-\delta_{ik}R_j-\delta_{jk}R_i+\frac{3 R_iR_j R_k}{R^2}\bigg]\, .
\end{align}
Thus, in the static limit, we recovered the displacement and displacement gradient
fields of the Kelvin problem, which is concerned with a concentrated point force
in three-dimensional elastostatics.

\section{The elastodynamic Li\'enard-Wiechert potentials and elastic fields 
of non-uniformly moving line forces}
We proceed to derive the two-dimensional Li\'enard-Wiechert potentials 
and the elastic fields of non-uniformly moving 
line forces.
We consider line forces moving non-uniformly and with time-dependent
magnitude. The line forces are parallel to the $x_3$-direction. 
In two dimensions, the in-plane line force is given by
\begin{align}
\label{F-2D}
F_{\alpha}=Q_\alpha(t)\, \delta(\BR(t))\, ,
\end{align}
and the anti-plane line force reads
\begin{align}
\label{F-z}
F_3=Q_3(t)\, \delta(\BR(t))\,,
\end{align}
where $\BR(t)=\rr-\Bs(t) \in \R^2$ and $\alpha=1,2$. 
The field variables are independent of the $x_3$-component.

If the material is infinitely extended and isotropic,  
the two-dimensional elastodynamic 
Green tensor of plane-strain reads~\citep{Eringen75,Kausel}
\begin{align}
\label{GT-2D}
G_{\alpha\beta}(\rr,t)&=\frac{1}{2\pi\rho }\, 
\Bigg\{
\frac{x_\alpha x_\beta}{r^4}\,
\bigg(
\frac{\big[2t^2-r^2/c^2_\TL\big]}{\sqrt{t^2-r^2/c^2_\TL}} 
\, H\big(t-r/c_\TL\big)
-\frac{\big[2t^2-r^2/c^2_\TT\big]}{\sqrt{t^2-r^2/c^2_\TT}}
\ H\big(t-r/c_\TT\big) 
\bigg)\nonumber\\
&\qquad\qquad
-\frac{\delta_{\alpha\beta}}{r^2}\, 
\bigg(
\sqrt{t^2-r^2/c^2_\TL}\, H\big(t-r/c_\TL\big)
-\frac{t^2}{\sqrt{t^2-r^2/c^2_\TT}}
\, H\big(t-r/c_\TT\big)
\bigg)\Bigg\}
\end{align}
and the elastodynamic Green tensor of anti-plane strain is given by
\begin{align}
\label{GT-zz}
G_{33}(\rr,t)=\frac{1}{2\pi\rho c_\TT^2}\, 
\frac{H\big(t-r/c_\TT\big)}{\sqrt{t^2-r^2/c^2_\TT}}\,,
\end{align}
where $H(.)$ denotes the Heaviside step function and $r=\sqrt{x_1^2+x_2^2}$.

If we substitute Eqs.~(\ref{GT-2D}) and (\ref{F-2D}) in Eq.~(\ref{u-M-L}) 
and perform the integration in $\rr'$, we find for the displacement field
\begin{align}
\label{u-2D}
u_{\alpha}(\rr,t)&=\frac{1}{2\pi\rho}
\bigg[
\int_{-\infty}^{t_{{\TL}}}  Q_\beta(t')
\bigg(\frac{ R_\alpha(t') R_\beta(t')}{R^4(t')}\, \frac{\bar{t}^2}{S_\TL(t')}
+\bigg(\frac{ R_\alpha(t') R_\beta(t')-\delta_{\alpha\beta}\, R^2(t')}{R^4(t')}
\bigg)S_\TL(t')\bigg)
\d t'\nonumber\\
&\qquad\quad
-\int_{-\infty}^{t_{{\TT}}} Q_\beta(t') 
\bigg(\frac{ R_\alpha(t') R_\beta(t')}{R^4(t')}\, S_\TT(t')
+\bigg(\frac{ R_\alpha(t') R_\beta(t')-\delta_{\alpha\beta}\, R^2(t')}{R^4(t')}
\bigg)
\frac{\bar{t}^2}{S_\TT(t')}\bigg)\d t'
\bigg]\, .
\end{align}
The notation here is 
\begin{align}
\label{Not-eg}
\bar{t}=t-t'\, , \qquad
&S^2_{\TT}(t')=\bar{t}^2-\frac{{R}^2(t')}{c^2_{\TT}}\, ,
\qquad
S^2_{\TL}(t')=\bar{t}^2-\frac{{R}^2(t')}{c^2_{\TL}}\, .
\end{align}
The two retarded times $t_\TT$ and $t_\TL$
are the roots of $S_\TT^2(t')=0$ and $S_\TL^2(t')=0$, respectively,
which are less than $t$.
The solving of the conditions $S_\TT^2(t')=0$ and $S_\TL^2(t')=0$
is non-trivial for a general motion and can be complicated.
For a subsonic motion, the solutions for the retarded times
$t_\TT$ and $t_\TL$ are unique.
The corresponding elastic distortion and velocity of a non-uniformly moving
line force are
\begin{align}
\label{B-2D}
\beta_{\alpha\gamma}(\rr,t)&=\frac{1}{2\pi\rho}\,
\pd_\gamma \bigg[
\int_{-\infty}^{t_{{\TL}}}  Q_\beta(t')
\bigg(\frac{R_\alpha(t') R_\beta(t')}{R^4(t')}\, \frac{\bar{t}^2}{S_\TL(t')}
+\bigg(\frac{R_\alpha(t') R_\beta(t')-\delta_{\alpha\beta}\, R^2(t')}{R^4(t')}\bigg)S_\TL(t')\bigg)
\d t'\nonumber\\
&\qquad\quad
-\int_{-\infty}^{t_{{\TT}}} Q_\beta(t') 
\bigg(\frac{R_\alpha(t') R_\beta(t')}{R^4(t')}\, S_\TT(t')
+\bigg(\frac{R_\alpha(t') R_\beta(t')-\delta_{\alpha\beta}\, R^2(t')}{R^4(t')}\bigg)
\frac{\bar{t}^2}{S_\TT(t')}\bigg)\d t'
\bigg]\, 
\end{align}
and 
\begin{align}
\label{v-2D}
v_{\alpha}(\rr,t)&=\frac{1}{2\pi\rho}\,
\pd_t \bigg[
\int_{-\infty}^{t_{{\TL}}}  Q_\beta(t')
\bigg(\frac{R_\alpha(t') R_\beta(t')}{R^4(t')}\, \frac{\bar{t}^2}{S_\TL(t')}
+\bigg(\frac{R_\alpha(t') R_\beta(t')-\delta_{\alpha\beta}\, R^2(t')}{R^4(t')}
\bigg)S_\TL(t')\bigg)
\d t'\nonumber\\
&\qquad\quad
-\int_{-\infty}^{t_{{\TT}}} Q_\beta(t') 
\bigg(\frac{R_\alpha(t') R_\beta(t')}{R^4(t')}\, S_\TT(t')
+\bigg(\frac{R_\alpha(t') R_\beta(t')-\delta_{\alpha\beta}\, R^2(t')}{R^4(t')}\bigg)
\frac{\bar{t}^2}{S_\TT(t')}\bigg)\d t'
\bigg]\, .
\end{align}

Now we consider the anti-plane line force.
If we substitute Eqs.~(\ref{GT-zz}) and (\ref{F-z}) in Eq.~(\ref{u-M-L}) 
and perform the integration in $\rr'$, we find for the displacement field
of a line load of body forces pointing in the $x_3$-direction
\begin{align}
\label{u-z}
u_{3}(\rr,t)=\frac{1}{2\pi\rho c_\TT^2}
\int_{-\infty}^{t_{{\TT}}}
\frac{Q_3(t')}{S_\TT(t')}\,\d t'\, .
\end{align}
The corresponding elastic distortion and velocity fields
are
\begin{align}
\label{B-z}
\beta_{3\alpha}(\rr,t)=\frac{1}{2\pi\rho c_\TT^2}\,
\pd_\alpha \int_{-\infty}^{t_{{\TT}}}
\frac{Q_3(t')}{S_\TT(t')}\,\d t'\, ,\qquad
v_{3}(\rr,t)=\frac{1}{2\pi\rho c_\TT^2}\,
\pd_t \int_{-\infty}^{t_{{\TT}}}
\frac{Q_3(t')}{S_\TT(t')}\,\d t'\, .
\end{align}

The displacement fields of non-uniformly moving line forces 
play the role of the two-dimensional elastodynamical Li\'enard-Wiechert potentials.
The two-dimensional Li\'enard-Wiechert potentials~(\ref{u-2D}) and (\ref{u-z})
are time-integrals over the history of the motion and, thus,
they are characterized by an afterglow. 
Such two-dimensional wave motion possesses a `tail' characteristic for
the so-called diffusion of waves.
For that reason  line forces are haunted by their past.
Nevertheless, for the two-dimensional 
Li\'enard-Wiechert potentials~(\ref{u-2D})
and (\ref{u-z}) 
we are left to evaluate time-integrals of considerable complexity, which only 
in some simple cases yield results of elementary functions in a closed form.
Also, the calculation of the retarded times is not a trivial task.
The mathematical complexity of the integrals~(\ref{u-2D})--(\ref{B-z}) is the same 
as of the integral expressions of non-uniformly moving straight dislocations given
by~\citet{Lardner} and \citet{Lazar2011}.

If we put $s(t')=0$ in Eqs.~(\ref{u-2D}) and (\ref{u-z}),
we recover the displacement fields of concentrated line forces
given by~\citet{deHoop58} and \citet{Achenbach}.
For the in-plane line force, \citet{deHoop58} and \citet{Achenbach}
used a more complicated but equivalent representation of the 
Green tensor~(\ref{GT-2D}).

\section{Conclusion}
Exact analytical solutions of the displacement and of elastic fields 
have been calculated for 
point and line forces moving non-uniformly in an unbounded, elastic, isotropic body. 
We have investigated the subsonic motion ($|\BV|<c_\TT$).
We have shown that the displacements can be interpreted as the 
elastodynamic Li\'enard-Wiechert potentials caused by body forces. 
For a point force, we calculated explicitly the radiation parts of the elastic
fields.
We have proven that our solution of a non-uniformly moving
point force is the correct generalization of 
the Stokes solution and of the solution of the Kelvin problem.
In the case of line forces the Li\'enard-Wiechert potentials
are given in the form of time-integral representations,
which cannot be further simplified for the general non-uniform motion.

\section*{Acknowledgement}
The author gratefully acknowledges the grants of the 
Deutsche Forschungsgemeinschaft (Grant Nos. La1974/2-1, La1974/3-1). 
The author wishes to thank Dr. Eleni Agiasofitou 
for her valuable comments and suggestions to improve the paper.
In addition, the author wants to thank an anonymous reviewer 
for constructive comments helping to prepare the present revised version.

\end{document}